\documentclass[runningheads,a4paper]{llncs}
\usepackage{etex}
\reserveinserts{28}
\pdfoutput=1

\usepackage[linesnumbered,lined,boxed,commentsnumbered,ruled]{algorithm2e}

\SetAlFnt{\small}
\SetAlCapFnt{\small}
\SetAlCapNameFnt{\small}
\SetAlCapHSkip{0pt}
\IncMargin{-\parindent}
\usepackage{ textcomp }
\usepackage{ comment }
\usepackage[color=yellow!60,textsize=footnotesize,obeyDraft,draft]{todonotes}
\usepackage{colortbl}
\usepackage{booktabs}
\usepackage{capt-of}
\usepackage{makecell}
\usepackage{textcomp }
\usepackage{multirow}
\usepackage{listings}
\usepackage{graphicx}
\usepackage{colortbl}
\usepackage{booktabs}
\usepackage{amsmath}
\usepackage[font=footnotesize,labelfont=bf]{subcaption}
\usepackage[hidelinks]{hyperref}
\usepackage[capitalize]{cleveref}
\usepackage{tikz}
\usepackage{pgfplots}
\usepackage[para,online,flushleft]{threeparttable}
\usepackage{multirow}
\usepackage{wasysym}
\usepackage{amssymb}
\usepackage{enumitem}
\usetikzlibrary{arrows}
\usepackage{subcaption}
\usepackage{breakcites}
\usepackage{url}
\newcounter{mnotei}
\newcolumntype{L}[1]{>{\raggedright\let\newline\\\arraybackslash\hspace{0pt}}m{#1}}
\newcolumntype{C}[1]{>{\centering\let\newline\\\arraybackslash\hspace{0pt}}m{#1}}
\newcolumntype{R}[1]{>{\raggedleft\let\newline\\\arraybackslash\hspace{0pt}}m{#1}}

\newcommand{\includegraphicsmaybe}[2]{
    \IfFileExists{#2}{\includegraphics[#1]{#2}}{
    \detokenize{File #2 is missing, maybe you need to run plots.py?}
}}

\begin{document}
\mainmatter

\title{Less is More: Exploiting the Standard Compiler Optimization Levels for Better Performance and Energy Consumption}

\titlerunning{Less is More: Exploiting the Standard Compiler Optimization Levels}

\author{Kyriakos Georgiou\inst{1}, Craig Blackmore\inst{1}, Samuel Xavier-de-Souza\inst{2}, Kerstin Eder\inst{1}}

\authorrunning{K. Georgiou et al.}
\institute{University of Bristol, UK \and Universidade Federal do Rio Grande do Norte, Brazil}
\tocauthor{Authors' Instructions}
\maketitle

\makeatletter
\renewcommand\subsubsection{\@startsection{subsubsection}{3}{\z@}%
                       {-18\p@ \@plus -4\p@ \@minus -4\p@}%
                       {4\p@ \@plus 2\p@ \@minus 2\p@}%
                       {\normalfont\normalsize\bfseries\boldmath
                        \rightskip=\z@ \@plus 8em\pretolerance=10000 }}
\makeatother

\begin{abstract}
This paper presents the interesting observation that by performing fewer of the optimizations available in a standard compiler optimization level such as -O2, while preserving their original ordering, significant savings can be achieved in both execution time and energy consumption. This observation has been validated on two embedded processors, namely the ARM Cortex-M0 and the ARM Cortex-M3, using two different versions of the LLVM compilation framework; v3.8 and v5.0. Experimental evaluation with 71 embedded benchmarks demonstrated performance gains for at least half of the benchmarks for both processors. An average execution time reduction of 2.4\% and 5.3\% was achieved across all the benchmarks for the Cortex-M0 and Cortex-M3 processors, respectively, with execution time improvements ranging from 1\% up to 90\% over the -O2. The savings that can be achieved are in the same range as what can be achieved by the state-of-the-art compilation approaches that use iterative compilation or machine learning to select flags or to determine phase orderings that result in more efficient code. In contrast to these time consuming and expensive to apply techniques, our approach only needs to test a limited number of optimization configurations, less than 64, to obtain similar or even better savings. Furthermore, our approach can support multi-criteria optimization as it targets execution time, energy consumption and code size at the same time. 
\end{abstract}

\section{Introduction}

Compilers were introduced to abstract away the ever-increasing complexity of hardware and  improve software development productivity. At the same time, compiler developers face a hard challenge: producing optimized code. A modern compiler supports a large number of architectures and programming languages and it is used for a vast diversity of applications. Thus, tuning the compiler optimizations to perform well across all possible applications is impractical. The task is even harder as compilers need to adapt to rapid advancements in hardware and programming languages.

Modern compilers adopted two main practices to mitigate the problem and find a good balance between the effort needed to develop compilers and their effectiveness in optimizing code. The first approach is the splitting of the compilation process into distinct phases. Modern compilers such as those based on the LLVM compilation framework~\cite{Lattner:MSThesis02}, allow for a common optimizer that can be used by any architecture and programming language. This is made possible by the use of an Intermediate Representation (IR) language on which  optimizations are applied. Then a front-end framework is provided to allow programming languages to be translated into the IR, and a back-end framework exists that allows the IR to be translated into specific instruction set architectures (ISA). Therefore, to take advantage of the common optimizer one only needs to create a new front-end for a programming language and a new back-end for an architecture. 

The second practice is the use of standard optimization levels, typically \mbox{-O0}, -O1, -O2, -O3 and -Os. Most modern compilers have a large number of transformations exposed to software developers via compiler flags; for example, the LLVM's optimizer has 56 documented transformations~\cite{LLVM:passes2018}. There are two major challenges a software developer faces while using compilers. First, to select the right set of transformations, and second to order the chosen transformations in a meaningful way, also called the compiler phase-ordering problem. The common objective is to achieve the best resource usage based on the application's requirements. 
To address this, each standard optimization level offers a predefined sequence of optimizations, which are proven to perform well based on a number of micro-benchmarks and a range of architectures. For example, for the LLVM compilation framework, starting from the -O0 level, which has no optimizations enabled, and moving to -O3, each level offers more aggressive optimizations with the main focus being performance, while -Os is focused on optimizing code size. Code size is critical for embedded applications with a limited amount of memory available. Furthermore, the optimization sequences defined for each level encapsulate the accumulated empirical knowledge of compiler engineers over the years. For example, some optimizations depend on other code transformations being applied first, and some optimizations offer more opportunities for other optimizations. Note that a code transformation is not necessarily an optimization, but instead, it can facilitate an IR structure which enables the application of other optimizations. Thus, a code transformation does not always lead to better performance. 
  
Although standard optimization levels are a good starting point, they are far from optimal in many cases, depending on the application and architecture used. An optimization configuration is a sequence of ordered flags. Due to the huge number of possible flag combinations and their possible orderings, it is impractical to explore the whole optimization-configuration space. Thus, finding optimal optimization configurations is still an open challenge. To tackle this issue, iterative compilation and machine-learning techniques have been used to find good optimization sequences by exploiting only a fraction of the optimization space~\cite{2018arXiv180104405A}. Techniques involving iterative compilation are expensive since typically a large amount of optimization configurations, in the order of hundreds to thousands, need to be exercised before reaching any performance gains over standard optimization levels. On the other hand, machine learning approaches require a large training phase and are hardly portable across compilers and architectures.

This paper takes a different approach. Instead of trying to explore a fraction of the whole optimization space, we are focusing on exploiting the existing optimization levels. For example, using the optimization flags included in the -O2 optimization level as a starting point, a new optimization configuration is generated each time by removing the last transformation flag of the current optimization configuration. In this way, each new configuration is a subsequence of the -O2 configuration, that preserves the ordering of flags  in the original optimization level. Thus, each new optimization configuration stops the optimization earlier than the previously generated configuration did. This approach aims to preserve the empirical knowledge built into the ordering of flags for the standard optimization levels. The advantages of using this technique are:
\begin{itemize}
\item The architecture and the compiler are treated as a black box, and thus, this technique is easy to port across different compilers or versions of the same compiler, and different architectures. To demonstrate this we applied our approach to two embedded architectures (Arm Cortex-M0 and Cortex-M3) and two versions of the LLVM compilation framework (v3.8 and v5.0);
\item An expensive training phase similar to the ones needed by the machine learning approaches is not required;
\item The empirical knowledge built into the existing optimization levels by the compiler engineers is being preserved;
\item In contrast to machine-learning approaches and random iterative compilation~\cite{bodin:inria-00475919}, which permit reordering transformation passes, our technique retains the original order of the transformation passes. Reordering can break the compilation or create a malfunctioning executable;
\item In contrast to the majority of machine-learning approaches, which are often opaque, our technique provides valuable insights to the software engineer on how each optimization flag affects the resource of interest;
\item  Because energy consumption, execution time and code size of each optimization configuration are being monitored during compilation, multi-criteria optimizations are possible without needing to train a new model for each resource.
\end{itemize}

Our experimental evaluation demonstrates an average of 2.4\% and 5.3\% execution time improvement for the Cortex-M0 and Cortex-M3 processors, respectively. Similar savings were achieved for energy consumption. These results are in the range of what existing complicated machine learning or time consuming iterative compilation approaches can offer on the same embedded processors~\cite{blackmore2015,Pallister2015}.

The rest of the paper is organized as follows. \Cref{sec:comp_and_analysis} gives an overview of the compilation and analysis methodology used. Our experimental evaluation methodology, benchmarks and results are presented and discussed in \cref{sec:results}. \Cref{sec:related_work} critically reviews previous work related to ours. Finally, \Cref{sec:conc_future} concludes the paper and outlines opportunities for future work.

\section{Compilation and Analysis}
\label{sec:comp_and_analysis}

As the primary focus of this work is deeply embedded systems, we demonstrate the portability of our technique across different architectures by exploring two of the most popular embedded processors: the Arm Cortex-M0~\cite{Cortex_M0} and the Arm Cortex-M3~\cite{Cortex_M3}. Although the two architectures belong to the same family, they have significant differences in terms of performance and power consumption characteristics~\cite{Cortex_M_processors}. The technique treats an architecture as a black box as no resource models are required e.g.\ energy-consumption or execution-time models. Instead, execution time and energy consumption physical measurements are used to assess the effectiveness of a new optimization configuration on a program.

For demonstrating the portability of the technique across different compiler versions, the analysis for the Cortex-M0 processor was performed using the LLVM compilation framework v3.8., and for the Cortex-M3 using the LLVM compilation framework v5.0. The technique treats the compiler as a black box since it only uses the compilation framework to exercise the different optimization-configuration scenarios, extracted from a predefined optimization level, on a particular program. In contrast, machine-learning-based techniques typically require a heavy training phase for each new compiler version or when a new optimization flag is introduced~\cite{Ashouri:2017, blackmore2015}.

\begin{figure}[!t]
\centering
\includegraphics[trim={0.1cm 15.2cm 17.5cm 0cm}, scale=1 ,clip]{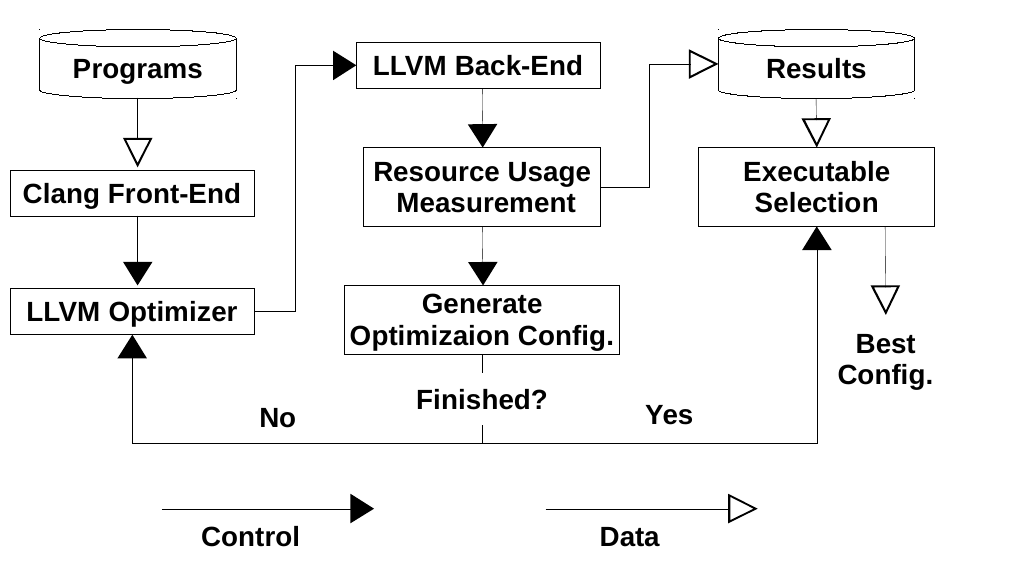}
\caption{Compilation and evaluation process.}
\label{fig_evaluation_process}
\end{figure}

\Cref{fig_evaluation_process} demonstrates the process used to evaluate the effectiveness of the different optimization configurations explored. Each configuration is a set of ordered flags used to drive the analysis and transformation passes by the LLVM optimizer. An analysis pass can identify properties and expose optimization opportunities that can later be used by transformation passes to perform optimizations. A standard optimization level (-O1, -O2, -O3, -Os, -Oz) can be selected as the starting point. Each optimization level represents a list of optimization flags which have a predefined order. Their order influences the order in which the transformation / optimization and analysis passes will be applied to the code under compilation. A new flag configuration is obtained by excluding the last transformation flag from the current list of flags. 
Then the new optimization configuration is being applied to the unoptimized intermediate representation (IR) of the program, obtained from the Clang front-end. Note that the program's unoptimized IR only needs to be generated once by the Clang front-end; it can then be used throughout the exploration process thus saving compilation time. The optimized IR is then passed to the LLVM back-end and linker to generate the executable for the architecture under consideration. Note that both the back-end and linker are always called using the optimization level selected for exploration; in our case -O2. The executable's energy consumption, execution time and code size are measured and stored. The exploration process finishes when the current list of transformation flags is empty. This is equivalent to optimization level -O0, where no optimizations are applied by the optimizer. Then, depending on the resource requirements, the best flag configuration is selected.

There are two kinds of pass dependencies for the LLVM optimizer; explicit and implicit dependencies. An explicit dependency exists when a transformation pass requires an other analysis pass to execute first. In this case, the optimizer will automatically schedule the analysis pass if only the transformation pass was requested by the user. An implicit dependency exists when a transformation or analysis pass is designed to work after another transformation instead of an analysis pass. In this case, the optimizer will not schedule the pass automatically, instead the user must manually add the passes in the correct order to be executed either using the \emph{opt} tool or the \emph{pass manager}. The \emph{pass manager} is the LLVM built-in mechanism for scheduling passes and handling their dependencies. If a pass is requested but its dependencies have not been requested in the correct order, then the specified pass will be automatically skipped by the optimizer. For the predefined optimization levels, the implicit dependencies are predefined in the \emph{pass manager}.

To extract the list of transformation and analysis passes, their ordering, and their dependencies for a predefined level of optimization, we use the argument "-debug-pass=Structure" with the \emph{opt} tool (the LLVM optimizer). This information is passed to our flag-selection process, which, to extract a new configuration, simply eliminates the last optimization flag applied. This ensures that all the implicit dependencies for the remaining passes in the new configuration are still in place. Thus, the knowledge built into the predefined optimization levels about effective pass orderings is preserved in the newly generated optimization configurations. What we are actually questioning is whether the pass scheduling in the predefined-optimization levels is a good choice. In other words, can stopping the optimizations at an earlier point yield more optimal code for a specific program and architecture? 

The BEEBS benchmark suite~\cite{DBLP:journals/corr/PallisterHB13} was used for evaluation. BEEBS is design for assessing the energy consumption of embedded processors. The resource usage estimation process retrieves the execution time, energy consumption and code size for each executable generated. The code size can be retrieved by examining the size of the executable. The execution time and energy consumption is being measured using the MAGEEC board~\cite{MAGEEC_board} together with the pyenergy~\cite{pyenergy} firmware and host-side software. 
The BEEBS benchmark suite utilizes this energy measurement framework and allows for triggering the begin and the end of the execution of a benchmark. Thus, energy measurements are reported only during a benchmark's execution. Energy consumption, execution time and average power dissipation are reported back to the host. The MAGEEC board supports a sampling rate of up to six million samples per second. A calibration process was needed prior to measurement to determine the number of times a benchmark should be executed in a loop while measuring to obtain an adequate number of measurements. This ensured the collection of reliable energy values for each benchmark. Finally, the BEEBS benchmark suite has a built-in self-test mechanism that flags up when a generated executable is invalid, i.e.\ it does not provide the expected results. Standard optimization levels shipped with each new version of a compiler are typically heavily tested to ensure the production of functionally correct executables. In our case, using optimization configurations that are subsequences of the standard optimization levels increases the chance of generating valid executables. In fact, all the executables we tested passed the BEEBS validation.

\begin{figure}[!htp]
    \centering
    \begin{subfigure}{\textwidth}
    \centering
      \includegraphics[trim=0.25cm 0.25cm 0.25cm 0.3cm,clip=true,width=\linewidth]{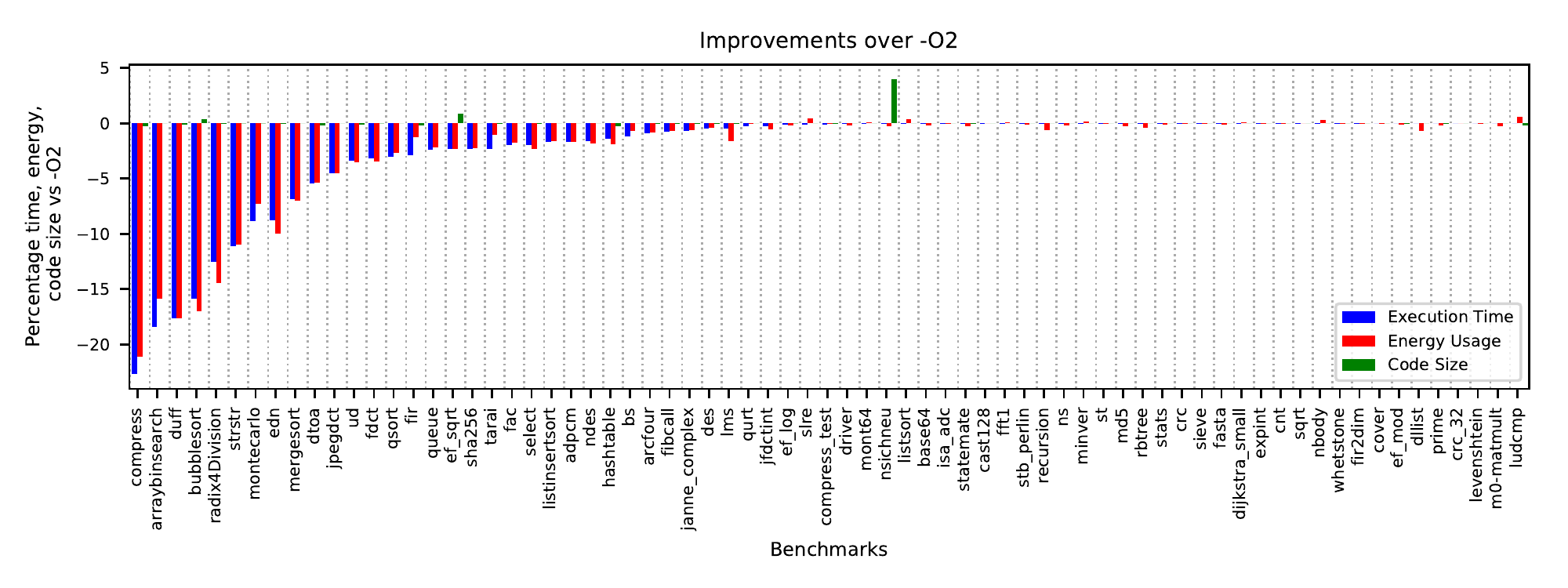}
      \subcaption{Results for the Cortex-M0 processor and the LLVM v3.8 compilation framework.}
      \label{subfig:CortexM0_total_result}
    \end{subfigure} 
    \begin{subfigure}{\textwidth}
    \centering
        \includegraphics[trim=0.25cm 0.25cm 0.25cm 0.3cm,clip=true,width=\linewidth]{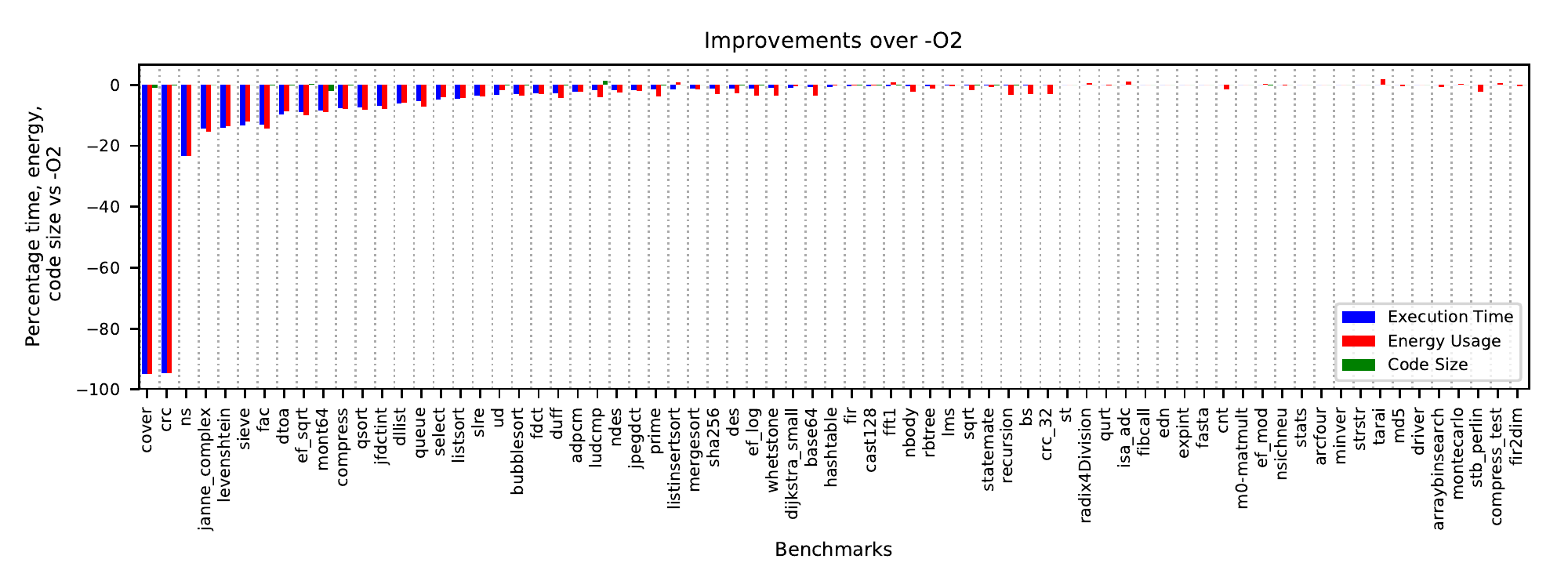}
      \subcaption{Results for the Cortex-M3 processor and the LLVM v5.0 compilation framework.}
      \label{subfig:CortexM3_total_result}
    \end{subfigure}
    \caption{Best achieved execution-time improvements over the standard optimization level -O2. For the best execution-time optimization configuration, energy consumption and code size improvements are also given. A negative percentage represents a reduction of resource usage compared to -O2.}
    \label{fig:all_results_graphs}
\end{figure}

\section{Results and Discussion}
\label{sec:results}

For the evaluation of our approach, the same 71 benchmarks from the BEEBS \cite{DBLP:journals/corr/PallisterHB13} benchmark suite were used for both the Cortex-M0 and the Cortex-M3 processors. For each benchmark, \Cref{fig:all_results_graphs} (\Cref{subfig:CortexM0_total_result} for the Cortex-M0 and the LLVM v3.8 and \Cref{subfig:CortexM3_total_result} for the Cortex-M3 and the LLVM v5.0) demonstrates the biggest performance gains achieved by the proposed technique compared to the standard optimization level under investigation, -O2. 
In other words, this figure represents the resource usage results obtained by using the optimization configuration, among the configurations exercised by our technique, that achieves the best performance gains compared to -O2 for each benchmark. 
A negative percentage represents an improvement on a resource, e.g.\ a result of -20\% for execution time represents a 20\% reduction in the execution time obtained by the selected optimization configuration when compared to the execution time retrieved by -O2. The energy-consumption and code-size improvements are also given for the selected configurations. If two optimization configurations have the same performance gains, then energy consumption improvement is used as a second criterion and code size improvement as a third criterion to select the best optimization configuration. The selection criteria can be modified according to the resource requirements for a specific application. Moreover, a function can be introduced to further formalize the selection process when complex multi-objective optimization is required. 

\begin{figure}[!htp]
    \centering
    \begin{subfigure}{\textwidth}
    \centering
    \begin{subfigure}{\textwidth}
    \centering
      \includegraphics[trim=0.25cm 0.25cm 0.25cm 0.3cm,clip=true,width=\linewidth]{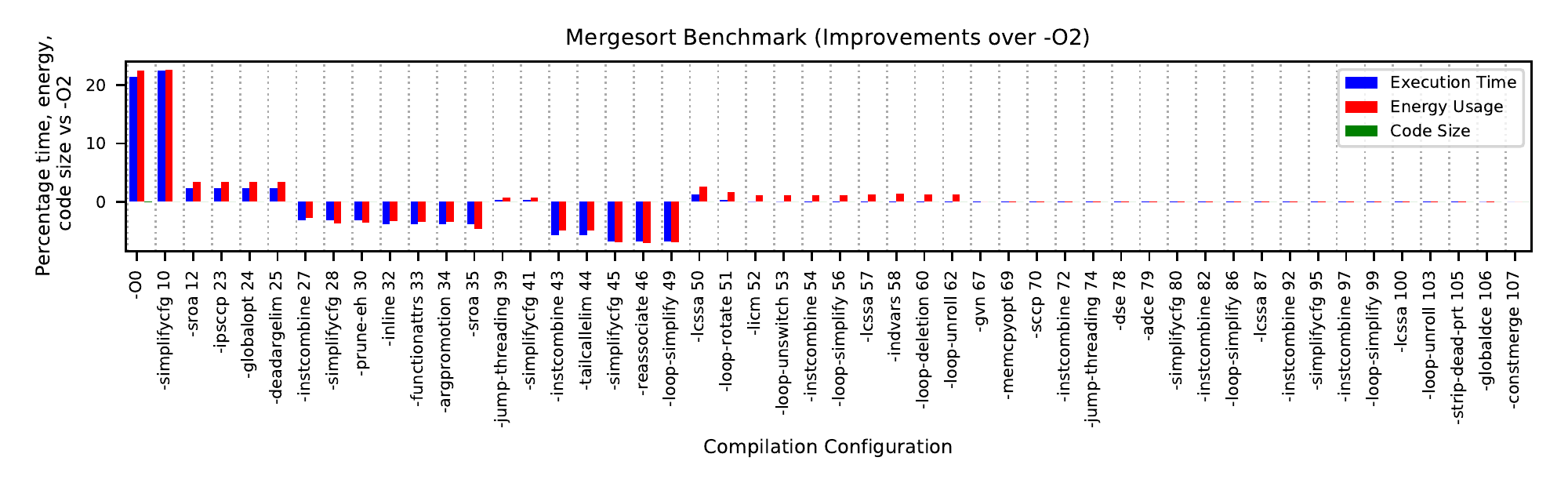}
    \end{subfigure} 
    \begin{subfigure}{\textwidth}
    \centering
        \includegraphics[trim=0.25cm 0.25cm 0.25cm 0.3cm,clip=true,width=\linewidth]{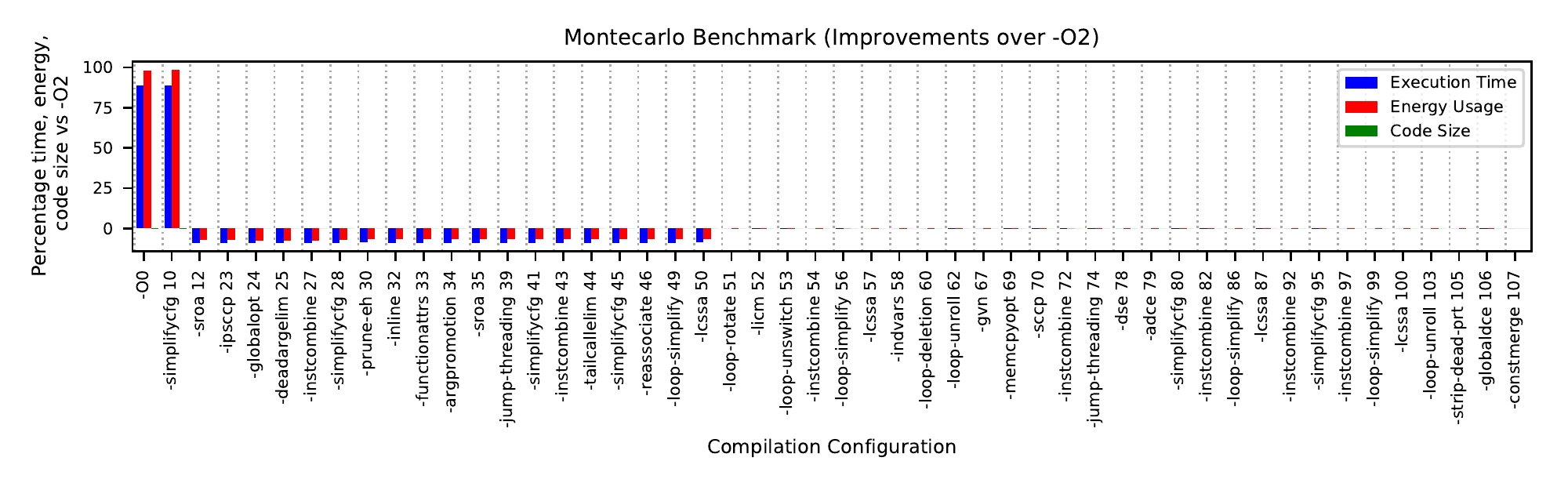}
    \end{subfigure}
      \subcaption{Compilation profiles for two of the benchmarks, using the Cortex-M0 processor and the LLVM v3.8 compilation framework.}
      \label{subfig:CortexM0_benchs_results}
    \end{subfigure}
    \begin{subfigure}{\textwidth}
    \centering
    \begin{subfigure}{\textwidth}
    \centering
      \includegraphics[trim=0.25cm 0.25cm 0.25cm 0.1cm,clip=true,width=\linewidth]{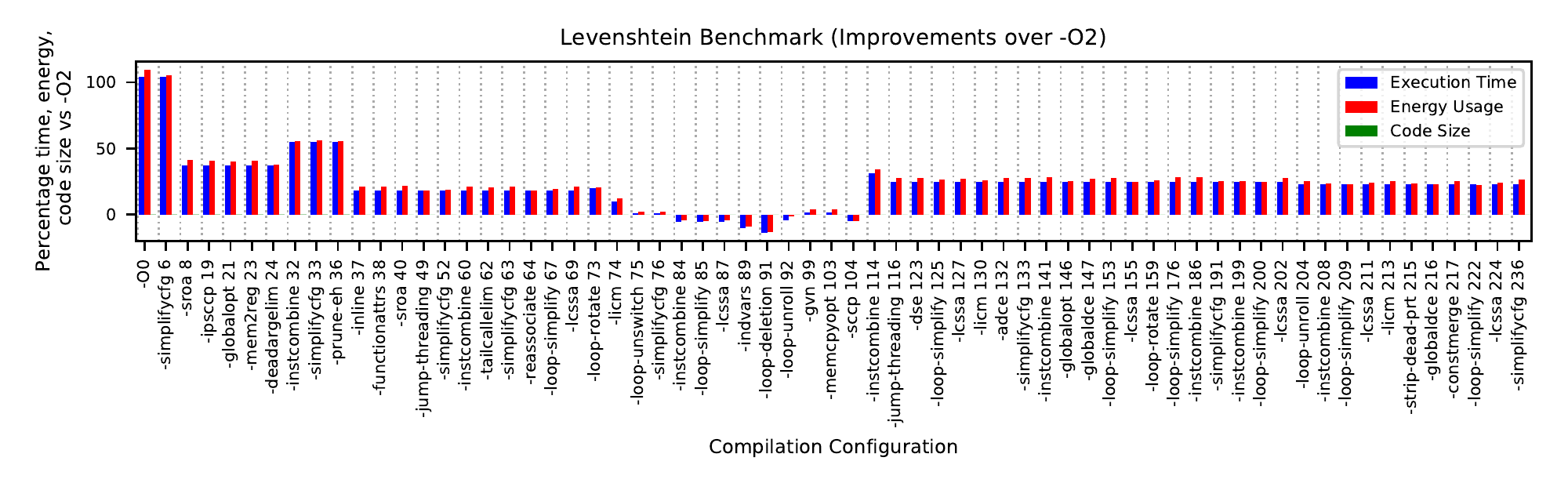}
    \end{subfigure} 
    \begin{subfigure}{\textwidth}
    \centering
        \includegraphics[trim=0.25cm 0.25cm 0.25cm 0.1cm,clip=true,width=\linewidth]{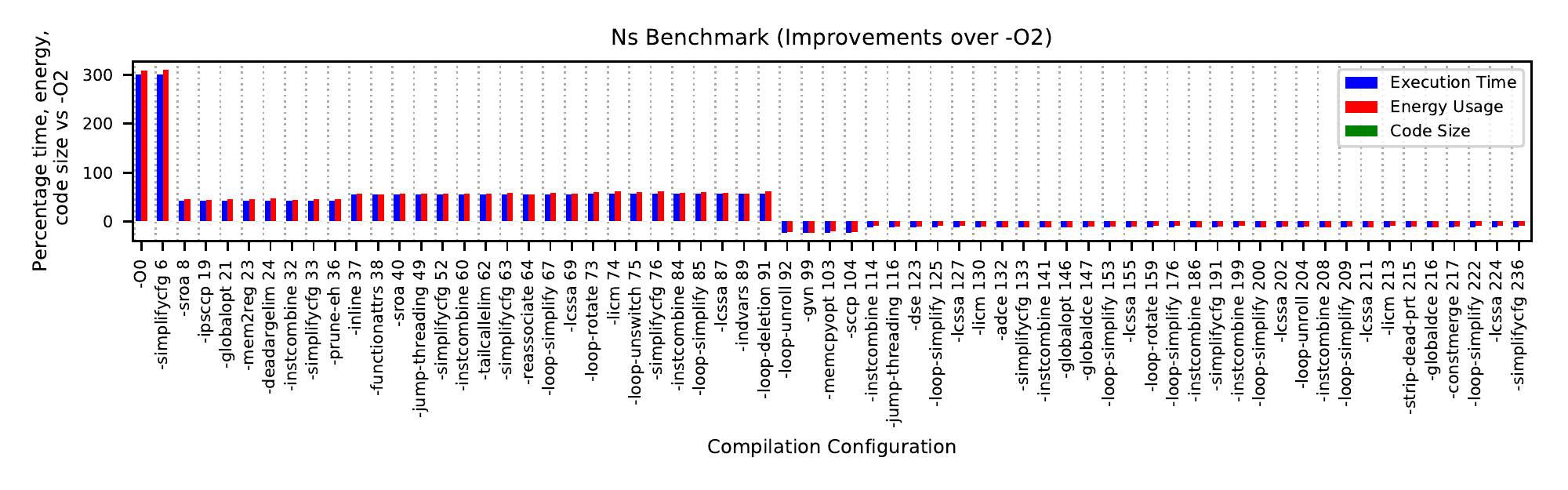}
    \end{subfigure}
    \subcaption{Compilation profiles for two of the benchmarks, using the Cortex-M3 processor and the LLVM v5.0 compilation framework}
    \label{subfig:CortexM3_benchs_results}
    \end{subfigure}
    \caption{For each optimization configuration tested by the proposed technique the execution-time, energy-consumption and code-size improvements over -O2 are given. A negative percentage represents a reduction of resource usage compared to -O2. Each element of the horizontal axis has the name of the last flag applied and the total number of flags used. The configurations are incremental subsequences of the -O2, starting from -O0 and adding optimization flags till reaching the complete -O2 set of flags.}
    \label{fig:benchs_results}
\end{figure}

For the Cortex-M0 processor, we observed an average reduction in execution time of 2.5\%, with 29 out of the 71 benchmarks seeing execution time improvements over -O2 ranging from around 1\% to around 23\%. For the Cortex-M3 processor, we observed an average reduction in execution time of 5.3\%, with 38 out of the 71 benchmarks seeing execution time improvements over -O2 ranging from around 1\% to around 90\%. The energy consumption improvements were always closely related to the execution time improvements for both of the processors. This is expected due to the predictable nature of these deeply embedded processors. In contrast, there were no significant fluctuations in the code size between different optimization configurations. We anticipate that, if the -Os or -Oz optimization levels, which both aim to achieve smaller code size, had been used as a starting point for our exploration, then more variation would have been observed for code size.

As it can be seen from \Cref{subfig:CortexM0_total_result,subfig:CortexM3_total_result}, our optimization strategy performed significantly different for the two processors per benchmark. This can be caused by the different performance and power consumption characteristics of the two processors and/or the use of different compiler versions in each case. Furthermore, the technique performed better on the Cortex-M3 with the LLVM v5.0 compilation framework. This could be due to the compilation framework improvements from version 3.8 to version 5.0. Another possible reason might be that the -O2 optimization level for LLVM v5.0 includes more optimization flags than the LLVM v.3.8. The more flags in an optimization level, the more optimization configurations will be generated and exercised by our exploitation technique, and thus, more opportunities for execution-time, energy-consumption and code-size savings can be exposed.

\Cref{subfig:CortexM0_benchs_results,subfig:CortexM3_benchs_results} demonstrate the effect of each optimization configuration, exercised by our exploitation technique, on the three resources (execution time, energy consumption and code size), for two of the benchmarks for the Cortex-M0 and Cortex-M3 processors, respectively. Similar figures were obtained for all the 71 benchmarks and for both of the processors. Similarly to \Cref{fig:all_results_graphs}, a negative percentage represents an improvement on the resource compared to the one achieved by -O2. The horizontal axis of the figures shows the flag at which compilation stopped together with the total number of flags included up to that point. This represents an optimization configuration that is a subsequence of the -O2. For example, the best optimization configuration for all three resources for the \emph{Levenstein} benchmark (see top part of \Cref{subfig:CortexM3_benchs_results}) is achieved when the compilation stops at flag number 91, \emph{-loop-deletion}. This means that the optimization configuration includes the first 91 flags of the -O2 configuration with their original ordering preserved. The optimization configurations include both transformation and analysis passes. 

The number of optimization configurations exercised in each case depends on the number of transformation flags included in the -O2 level of the version of the LLVM optimizer used. Note that we are only considering the documented transformation passes~\cite{LLVM:passes2018}. For example, 50 and 64 different configurations are being tested in the case of the Cortex-M0 processor with the LLVM compilation framework v3.8, and the case of Cortex-M3 with the LLVM framework v5.0, respectively. Many of the transformation passes are repeated multiple times in a standard optimization level, but because of their different ordering, they have a different effect. Thus, we consider each repetition as an opportunity to create a new optimization configuration. Furthermore, note that more transformation passes exist in the LLVM optimizer, but typically, these are passes that have implicit dependencies on the documented passes. The methodology of creating a new optimization configuration explained in \Cref{sec:comp_and_analysis} ensures the preservation of all the implicit dependencies for each configuration. This is part of preserving the empirical knowledge of good interactions between transformations built into the predefined optimization levels and reusing it in the new configurations generated.

Typically, optimization approaches based on iterative compilation are extremely time consuming~\cite{Ashouri:2017}, since thousands of iterations are needed to reach levels of resource savings similar to the ones achieved by our approach. In our case the maximum number of iterations we had to apply were the 64 iterations for the Cortex-M3 processor. This makes our simple and inexpensive approach an attractive alternative, before moving to the more expensive approaches, such as iterative-compilation-based and machine-learning-based compilation techniques~\cite{2018arXiv180104405A,Asouri_thesis2016}.

By manually observing the compilation profiles obtained for all the benchmarks, similar to the ones demonstrated in \Cref{fig:benchs_results}, no common behavior patterns were detected, except that typically there is a significant improvement on the execution time and the energy consumption at the third optimization configuration. Future work will use clustering to see if programs can be grouped together based on their compilation profiles. This can be useful to identify optimization sequences that perform well for a particular type of program. Furthermore, the retrieved optimization profiles can also give valuable insights to compiler engineers and software developers on the effect of each optimization flag on a specific program and architecture. It is beyond the scope of this work to investigate these effects.

\section{Related Work}
\label{sec:related_work}

Iterative compilation has been proved an effective technique for tackling both the problems of choosing the right set of transformations and for ordering them to maximize their effectiveness~\cite{Ashouri:2017}. The technique is typically used to iterate over different sets of optimizations with the aim of satisfying an objective function. Usually, each iteration involves some feedback, such as profiling information, to evaluate the effectiveness of the tested configuration. In random iterative compilation~\cite{bodin:inria-00475919}, random optimization sequences are generated, ranging from hundreds to thousands, and then used to optimize a program. Random iterative compilation has been proved to provide significant performance gains over standard optimization levels. Thus, it has become a standard baseline metric for evaluating the effectiveness of machine-guided compilation approaches~\cite{Fursin2011,Ashouri:2017,blackmore2015}, where the goal is to achieve better performance gains with less exploration time. Due to the huge number of possible flag combinations and their possible orderings, it is impossible to explore a large fraction of the optimization space. To mitigate this problem, machine learning is used to drive iterative compilation~\cite{Agakov:2006,Ogilvie:2017,Cavazos:2007}. 

Based on either static code features \cite{Fursin2011} or profiling data \cite{Cavazos:2007}, such as performance counters, machine learning algorithms try to predict the best set of flags to apply to satisfy the objective function with as few iterations as possible. The techniques have proven to be effective in optimizing the resource usage, mainly execution-time, of programs on a specific architecture but generally suffer from a number of drawbacks. Typically, these techniques require a large training phase \cite{Ogilvie:2017} to create their predictive models. Furthermore, they are hardly portable across different compilers or versions of the same compiler and different architectures. Even if a single flag is introduced to the set of a compiler's existing flags the whole training phase has to be repeated. Moreover, extracting some of the metrics that these techniques depend on, such as static code features, might require a significant amount of engineering.

A recent work that is focused on mitigating the phase-ordering problem, \cite{Ashouri:2017}, divided the -O3 standard optimization flags of the LLVM compilation framework v3.8, into five subgroups using clustering. Then they used iterative compilation and machine learning techniques to select optimization configurations by reordering the subgroups. The approach demonstrated average performance speedup of 1.31. An interesting observation is that 79\% of the -O3 optimization flags were part of a single subgroup with a fixed ordering that is similar to that used in the -O3 configuration. This suggests that the ordering of flags in a predefined optimization level is a good starting point for further performance gains. Our results actually confirm this hypothesis for the processors under consideration.

Embedded applications typically have to meet strict timing, energy consumption, and code-size constraints \cite{Georgiou2017}. Hand-written optimized code is a complex task and requires extensive knowledge of architectures. Therefore, utilizing the compilers optimizations to achieve optimal resource usage is critical.

In an attempt to find better optimization configurations than the ones offered by the standard optimization levels, the authors in~\cite{blackmore2015} applied inductive logic programming (ILP) to predict compiler flags that minimize the execution time of software running on embedded systems. This was done by using ILP to learn logical rules that relate effective compiler flags to specific program features. For their experimental evaluation they used the GCC compiler,~\cite{GCC_compiler}, and the Arm Cortex-M3 architecture; the same architecture used by this paper. Their method was evaluated on 60 benchmarks selected from the BEEBS benchmark suite; the same used in this work. They were able to achieve an average reduction in execution time of 8\%, with about half of the benchmarks seeing performance improvements. The main drawback of their approach was the large training phase of their predictive model. For each benchmark, they needed to create and test 1000 optimization configurations. This resulted in about a week of training time. Furthermore, for their approach to be transferred to a new architecture, compiler or compiler version, or even to add a new optimization flag, the whole training phase has to be repeated from scratch. The same applies for applying their approach to resources other than execution time, such as energy consumption or code size. In contrast, our approach, for the same architecture and more benchmarks of the same benchmark suite, was able to achieve similar savings in execution time (average 5.3\%) by only testing 65 optimization configurations for each program. At the same time, our approach does not suffer from the portability issues faced by their technique. 

In~\cite{Pallister2015}, the authors used fractional factorial design (FFD) to explore the large optimization space ($2^{82}$ possible combinations for the GCC compiler used) and determine the effects of optimizations and optimization combinations. The resources under investigation were execution time and energy consumption. They tested their approach on five different embedded platforms including the Cortex-M0 and Cortex-M3, which are also used in this work. For their results to be statistically significant, they needed to exercise 2048 optimization configurations for each benchmark. Although they claimed that FFD was able to find optimization configurations that perform better than the standard optimization levels, they demonstrated this only on a couple of benchmarks. Again, this approach suffers from the same portability issues as~\cite{blackmore2015}.

In our work, to maximize the accuracy of our results, hardware measurements were used for both the execution time and energy consumption. Although, high accuracy is desirable, in many cases physical hardware measurements are difficult to deploy and use. Existing works demonstrated that energy modeling and estimation techniques could accurately estimate both execution time and energy consumption for embedded architectures similar to the ones used in this paper~\cite{Georgiou:2017,Grech:2015}. Such estimation techniques can replace the physical-hardware measurements used in our approach in order to make the proposed technique accessible to more software developers.

\section{Conclusion}
\label{sec:conc_future}

Finding optimal optimization configurations for a specific compiler, architecture, and program is an open challenge since the introduction of compilers. Standard optimization levels that are built-in to modern compilers, on average perform well on a range of architectures and programs and provide convenience to the software developer. Over the past years, iterative compilation and complex machine learning approaches have been exploited to yield optimization configurations that outperform these standard optimization levels. These techniques are typically expensive either due to their large training phases or the large number of configurations that they need to test. Moreover, they are hardly portable to new architectures and compilers.

In contrast, in this work an inexpensive and easily portable approach that generates and tests less than 64 optimization configurations proved able to achieve execution-time and energy-consumption savings in the same range of the ones achieved by state of the art machine learning and iterative compilation techniques~\cite{blackmore2015,Pallister2015,2018arXiv180104405A}. The effectiveness of this simple approach is attributed to the fact that we used subsequences of the optimization passes defined in the standard optimization levels, but stopped the optimizations at an earlier point than the standard optimization level under exploitation. This indicates that the accumulated empirical knowledge built into the standard optimization levels is a good starting point for creating optimization configurations that will perform better than the standard ones.

The approach is compiler and target independent. Thus, for its validation, two processors and two versions of the LLVM compiler framework were used; namely, the Arm Cortex-M0 with the LLVM v3.8 and the Arm Cortex-M3 with the LLVM v5.0. An average execution time reduction of 2.4\% and 5.3\% was achieved across all the benchmarks for the Cortex-M0 and Cortex-M3 processors, respectively, with at least half of the 71 benchmarks tested seeing performance and energy consumption improvements. Finally, our approach can support multi-criteria optimization as it targets execution time, energy consumption and code size at the same time.

In future work, clustering and other machine learning techniques can be applied on the compilation profiles retrieved by our exploitation approach (\Cref{fig:benchs_results}) to fine-tune the standard optimization levels of a compiler to perform better for a specific architecture. Furthermore, the technique is currently being evaluated on more complex  architectures, such as Intel's X-86.

\bibliographystyle{alphaurl}
\bibliography{bibliography}

\end{document}